\documentclass{epl}
\usepackage{epsfig}
\title{Self-diffusion of rod-like viruses in the nematic phase}
\author{M. P. Lettinga\inst{1} \and Edward Barry\inst{2} \and Zvonimir Dogic\inst{2}}
\institute{
  \inst{1} IFF, Institut Weiche Materie, Forschungszentrum J\"{u}lich,D-52425
J\"{u}lich, Germany\\
  \inst{2} Rowland Institute at Harvard, Harvard University, Cambridge MA 02142, USA\\
} 

\begin{document}

\maketitle

\begin{abstract}
We measure the self-diffusion of colloidal rod-like virus {\it fd}
in an isotropic and nematic phase. A low volume fraction of
viruses are labelled with a fluorescent dye and dissolved in a
background of unlabelled rods. The trajectories of individual rods
are visualized using fluorescence microscopy from which the
diffusion constant is extracted. The diffusion  parallel
($D_{\parallel}$) and perpendicular ($D_{\perp}$) to the nematic
director is measured. The ratio ($D_{\parallel}/D_{\perp}$)
increases monotonically with increasing virus concentration.
Crossing the isotropic-nematic phase boundary results in increase
of $D_{\parallel}$ and decrease of $D_{\perp}$ when compared to
the diffusion in the isotropic phase ($D_{\mbox{\scriptsize
iso}}$).
\end{abstract}

\section{Introduction}

Suspensions of semi-flexible polymers exhibit a variety of
dynamical phenomena, of great importance to both physics and
biology, that are still only partially understood. Advances over
the past decade include direct visual evidence for a
reptation-like diffusion of individual polymers in a highly
entangled isotropic solution and shape anisotropy of a single
polymer~\cite{Perkins94,Smith95,Kas96,Haber00}. If the
concentration of the polymers is increased, a suspension undergoes
a first order phase transition to a nematic phase, which has long
range orientational order but no long range positional order. As a
result of the broken orientational symmetry it is expected that
the diffusion of polymers in the nematic liquid crystals will be
drastically different from that in concentrated isotropic
solutions. While the static phase behavior of semi-flexible
nematic polymers is well understood in terms of the Onsager theory
and its extensions by Khoklov and
Semenov~\cite{Onsager49,Khokhlov81}, the dynamics of semi-flexible
polymers in the nematic phase is much less
explored~\cite{Dogic04}.

In this paper, we determine the concentration dependence of the
anisotropic diffusion of semi-flexible viruses in a nematic phase
and compare it to the diffusion in the isotropic phase.
Experimentally, the only data on the translational diffusion of
colloidal rods in the nematic phase was taken in a mixture of
labelled and unlabelled polydysperse boehmite rods using
fluorescence recovery after photobleaching
(FRAP)~\cite{Bruggen98}. Theoretically, molecular dynamics
simulations were performed on hard spherocylinder and ellipsoidal
systems from which the anisotropic diffusion data was
extracted~\cite{Allen90,Hess91,Lowen99}. The anisotropic diffusion
has also been studied in low molecular weight thermotropic liquid
crystals using NMR spectroscopy or inelastic scattering of
neutrons~\cite{deGennes93}.

Real space microscopy is a powerful method that can reveal
dynamics of colloidal and polymeric liquid systems that are
inaccessible to other traditional techniques~\cite{Kas96,Dogic04}.
We use digital microscopy to directly visualize the dynamics of
fluorescently labelled {\it fd} in a nematic background of the
unlabelled {\it fd}. The advantage of this method is an easy
interpretation of data and no need to obtain macroscopically
aligned monodomains in magnetic fields. The advantages of using
{\it fd} are its large contour length which can be easily
visualized with optical microscope and its phase behavior which
can be quantitatively described with the Onsager theory extended
to account for electrostatic repulsive interactions and
semi-flexibility~\cite{Tang95,Purdy03}. Viruses such as {\it fd}
and TMV have been used earlier to study the rod dynamics in the
isotropic phase~\cite{Cush02}.

\section{Experiment methods}

The physical characteristics of the bacteriophage {\it fd} are its
length L=880 nm, diameter D=6.6 nm, persistence length of 2200 nm
and a surface charge of 10 e$^-$/nm at pH 8.2~\cite{Dogic01}.
Bacteriophage {\it fd} suspension forms isotropic, cholesteric and
smectic phases with increasing
concentration~\cite{Dogic97,Dogic00c,Dogic01}. The free energy
difference between the cholesteric and nematic phase is very small
and locally the cholesteric phase is identical to nematic. We
expect that at short time scales the diffusion of the rods for
these two cases would be the same. Hereafter, we refer to the
liquid crystalline phase at intermediate concentration as a
nematic instead of a cholesteric.

The {\it fd} virus was prepared according to a standard biological
protocol using XL1-Blue strain of {\it E. coli} as the host
bacteria~\cite{Maniatis89}. The yields are approximately 50 mg of
{\it fd} per liter of infected bacteria and virus is typically
grown in 6 liter batches. Subsequently, the virus is purified by
repetitive centrifugation (108,000 g for 5 hours) and re-dispersed
in a 20 mM phosphate buffer at pH=7.5. First order
isotropic-nematic (I-N) phase transition for {\it fd} under these
conditions takes place at a rod concentration of 15.5 mg/ml.

Fluorescently labelled {\it fd} viruses were prepared by mixing 1
mg of {\it fd} with 1 mg of succinimidyl ester Alexa-488
(Molecular Probes) for 1 hour. The dye reacts with free amine
groups on the virus surface to form irreversible covalent bonds.
The reaction is carried out in small volume (100 $\mu$l, 100 mM
phosphate buffer, pH=8.0) to ensure a high degree of labelling.
Excess dye was removed by repeated centrifugation steps.
Absorbance spectroscopy indicates that there are approximately 300
dye molecules per each {\it fd} virus. Viruses labelled with
fluorescein isothiocynante, a dye very similar to Alexa 488,
exhibit the phase behavior identical to that of unlabelled virus.
Since liquid crystalline phase behavior is a sensitive test of
interaction potential, it is reasonable to assume that the
interaction potential between labelled viruses is very similar to
that between unlabeled viruses.

The samples were prepared by mixing one unit of anti-oxygen
solution (2 mg/ml glucose oxidase, 0.35 mg/ml catalase, 30 mg/ml
glucose and 5\% $\beta$-mercaptoethanol), one unit of a dilute
dispersion of Alexa 488 labelled viruses and eight units of the
concentrated {\it fd} virus suspension at the desired
concentration. Under these conditions the fluorescently labelled
viruses are relatively photostable and it is possible to
continuously observe rods for 3-5 minutes without significant
photobleaching. The ratio of labelled to unlabelled particles is
roughly kept at 1:30000. The samples were prepared by placing 4
$\mu$l of solution between a No 1.5 cover slip and coverslide. The
thickness of the samples is about 10 $\mu$m. Thin samples are
important to reduce the signal of out-of-focus particles. Samples
are equilibrated for half an hour, allowing flows to subside and
liquid crystalline defects to anneal. We have analyzed data at
various distances from the wall and have not been able to observe
a significant influence of wall on the diffusion of viruses.

For imaging we used an inverted Nikon TE-2000 microscope equipped
with $100\times$ 1.4 NA PlanApo oil immersion objective, a 100 W
mercury lamp and a fluorescence cube for Alexa 488 fluorescent
dye. The images where taken with a cooled CCD Camera (CoolSnap HQ,
Roper Scientific) set to an exposure time of 60 ms, running in a
overlap mode at a rate of 16 frames per second with $2\times2$
binning. The pixel size was 129 nm and the field of view was 89
$\mu$m $\times$ 66 $\mu$m. Typically there were around hundred
fluorescently labeled rods in the field of view. For each {\it fd}
concentration ten sequences of four hundred images were recorded.

\begin{figure}
\centerline{\epsfig{file=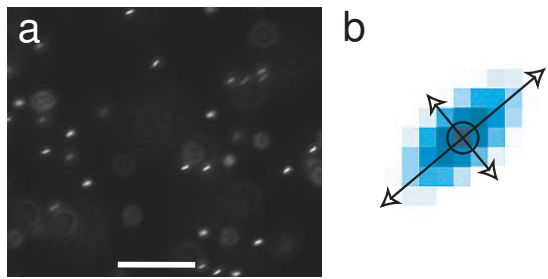,width=3in}}
\caption{\label{rodsinrods} (a) Image of fluorescently labelled
rods dissolved in a background nematic phase of unlabelled rods.
Scale bar is 5 $\mu$m. (b) Two-dimensional gaussian fit to a
individual rod. Arrows indicate the long and short axis. The
circle indicates the center of mass. From this fit it is possible
to obtain the orientation of an individual {\it fd} rod. Pixel
size is 129 nm.}
\end{figure}

\section{Analysis method}

Figure \ref{rodsinrods}a shows a typical image of fluorescently
labelled rods in a background nematic of unlabelled rods. Due to
limited spatial and temporal resolution of the optical microscope,
labelled {\it fd} appear as a slightly anisotropic rod, although
the actual aspect ratio is larger then 100. To measure the
anisotropic diffusion in the nematic phase, it is first necessary
to determine the nematic director which has to be uniform within a
field of view. Spatial distortion of the nematic would
significantly affects our results. The centers of mass and
orientation of rods are obtained sequentially. In a first step, a
smoothed image is used to identify the rods and obtain the
coordinates of its center of mass using image processing code
written in IDL~\cite{IDLnote}. Subsequently, a two dimensional
gaussian fit around a center of mass of each rod is performed
(Fig.~\ref{rodsinrods}b). From this fit the orientation of each
rod-like virus is obtained. This procedure is then repeated for a
sequence of images.

\begin{figure}
\centerline{\epsfig{file=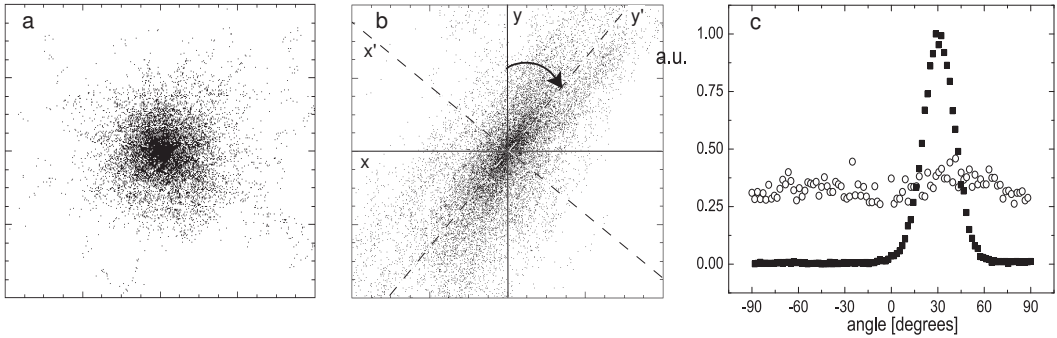,width=5.5in}}
\caption{\label{coorcloud} (a) A collection of trajectories of
fluorescently labelled virus particles in the isotropic phase. All
trajectories are translated so that the first point is located at
the origin. For clarity we only show the center of mass and not a
line connecting subsequent point in a particle trajectory. The
concentration of virus in this sample was 14 mg/ml. (b)
Anisotropic trajectories of the fluorescently labelled viruses
diffusing in the nematic phase. The concentration of the
background virus in this sample was 21 mg/ml. x' and y' indicate a
new lab-frame in which the director is aligned along the y' axis.
(c) The orientational distribution function obtained by plotting
the probability distribution function of the virus orientation for
isotropic (open circles) and nematic phase (full squares). The
orientation of the virus is obtained from two-dimensional gaussian
fits, an example of which is shown in Fig.~\ref{rodsinrods}b. The
angle of the nematic director obtained from (b) and (c) are almost
identical. }
\end{figure}

The length of a trajectory is usually limited to a few seconds,
after which the particles diffuse out of focus. In
Fig.~\ref{coorcloud}a and b we plot the trajectories of an
ensamble of particles for both isotropic and nematic sample. As
expected the trajectories in the isotropic phase are spherically
symmetric (Fig.~\ref{coorcloud}a) while those in the nematic phase
exhibit a pronounced anisotropy (Fig.~\ref{coorcloud}a). The
symmetric nature of the distribution indicates that there is no
drift or flow in our samples. We obtain the orientation of the
nematic director using two independent methods. One method is to
measure the main axis of the distribution shown in
Fig.~\ref{coorcloud}b. This procedure assumes that the diffusion
is largest along the nematic director. An alternative method is to
plot a histogram of rod orientations which are obtained from 2D
gaussian fits to each rod (Fig.~\ref{rodsinrods}b). The resulting
orientational distribution function (ODF) is shown in
Fig.~\ref{coorcloud}c. In principle, it should be possible to
obtain both the nematic director and order parameter from ODF
shown Fig.~\ref{coorcloud}c. We find that the order parameter
obtained in such a way is systematically higher then the order
parameter obtained from more reliable x-ray
experiment~\cite{Purdy03}. This is due to significant rotational
diffusion each rod undergoes during an exposure time of 60 ms.

The differences in the orientation of the nematic director
obtained using these two methods is always less then 5 degrees.
For the example shown in Figs. \ref{coorcloud}b and c, we obtain a
nematic director at an angle of $31.2\;^{\circ}$ while the peak of
the orientational distribution function lies at $30.2\;^{\circ}$.
The director can be ``placed'' along one of the two main axis by
rotating the lab-frame.

The diffusion coefficients of the rods parallel ($D_{\parallel}$)
and perpendicular ($D_{\perp}$) to the director are  calculated
from the x'- and y'-component of the mean square displacement.
When director lies along the y'-axis, $D_{\parallel}$ and
$D_{\perp}$ are given by:

\begin{eqnarray}\label{Eqmsd}
D_{\parallel}=\frac{1}{N}\frac{1}{2}\sum\{y'_i(t)-y'_i(0)\}^2\\
D_{\perp}=\frac{1}{N}\frac{1}{\sqrt{2}}\sum\{x'_i(t)-x'_i(0)\}^2,
\end{eqnarray}

\noindent where $N$ is the number of traced particles. To obtain
$D_{\perp}$, $D_{x}$ is multiplied with $\sqrt{2}$ since only one
component of the diffusion perpendicular to the director is
measured. The underlying assumption of our analysis is that the
nematic director is oriented in the field of view. For 10 $\mu$m
thin samples this is reasonable.

\section{Results and discussion}

\begin{figure}
\centerline{\epsfig{file=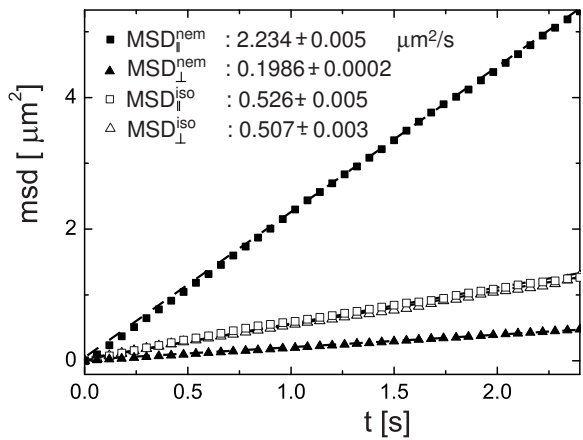,width=3in}}
\caption{\label{correlations} The mean square displacement of rods
along the director (full cubes) and perpendicular to the director
(full triangles) for a nematic sample at virus concentration of 21
mg/ml. The isotropic data are given by the open points and were
take just below I-N phase transition at virus concentration of 14
mg/ml. The diffusion along the director is significantly enhanced
when compared to the diffusion in the isotropic phase, while the
diffusion perpendicular to the director is significantly
suppressed. The mean square displacements shown in this figure are
measured from a single field of view.}
\end{figure}

Typical mean square displacements (MSD) are shown in Fig.
\ref{correlations} for  samples in an isotropic and nematic phase.
On average the mean square displacement was linear over fifty
frames in the nematic phase, but only about twenty five frames in
the isotropic phase. The diffusion perpendicular to the director
is slower in the nematic phase as compared to the isotropic phase.
Therefore in the nematic phase, the particles stay longer in focus
and can be tracked for a longer time. Since the MSD is linear over
the entire time range and displacements are up to a few times the
particle length, we are measuring pure long-time self-diffusion.
Visual inspection of the trajectories in the concentrated
isotropic phase, just below I-N coexistence shows no
characteristics of the reptation observed in suspensions of long
DNA fragments or actin filaments~\cite{Smith95,Kas96}. This points
to the fact that {\it fd} is very weakly entangled in a
concentrated isotropic suspension. This is in agreement with
recent microrheology measurements of {\it fd}
suspensions~\cite{Addas04}. We note that MSD's obtained from few
hundred trajectories within a single field of view are very
accurate. However, if we move to another region of the sample
sample we obtain MSD with slightly different slope. This leads to
conclusion that the largest source of error in measuring the
anisotropic diffusion coefficient is the uniformity of the nematic
director within the field of view.

\begin{figure}
\centerline{\epsfig{file=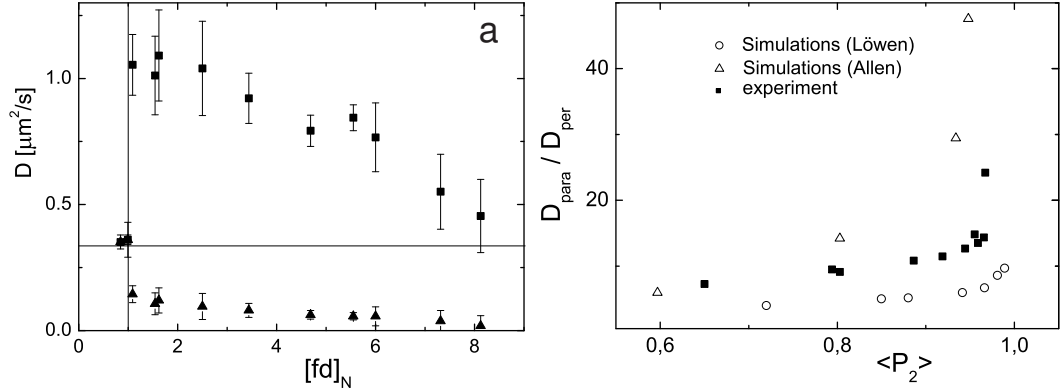,width=5.5in}}
\caption{\label{diffvscon} (a) The concentration dependence of the
translational diffusion parallel to ($D_{\parallel}$) and
perpendicular to ($D_{\perp}$) the nematic director are indicated
by squares and triangles respectively. The nematic phase in
coexistence with the isotropic phase occurs at $c_{fd}$=15.5 mg/ml
and is indicated by a vertical line. The x-axis is rescaled so
that I-N transition takes place at $[fd]_N$=1. (b) The plot of the
dimensionless ratio of the parallel to perpendicular diffusion
constant $D_{\parallel}$/$D_{perp}$ as a function of the nematic
order parameter. The concentration dependence of the nematic order
parameter is taken from ref.~\protect\cite{Purdy03}. Open
triangles are data for hard spherocylinders with aspect ratio of
10 taken from ref.~\protect\cite{Lowen99}  while open circles are
data for ellipsoids with aspect ratio 10 taken
from~\protect\cite{Allen90}}
\end{figure}

The concentration dependence of the anisotropic diffusion
constants is shown in Fig.~\ref{diffvscon}a. The nematic phase
melts into a isotropic phase at low concentrations and freezez
into a smectic phase at high concentrations. We made an attempt to
measure the diffusion of rods in the smectic phase, but have not
seen any appreciable diffusion on optical length scales over a
time period of minutes. The most strinking feature of our data is
a strong discontinuity in the behavior of the diffusion constant
at the I-N phase transition. Compared to diffusions in isotropic
case $D_{\mbox{\scriptsize iso}}$, $D_{\parallel}$ is larger by a
factor of four, while $D_{\perp}$ is smaller by a factor of two.
The concentration dependence of $D_{\parallel}$ and $D_{\perp}$
exhibit different behavior. With increasing concentration, for
$D_{\parallel}$ we measure an initial plateau, which is followed
by a broad region where the diffusion rate decreases
monotonically. $D_{\perp}$, however, shows a monotonic decrease of
the diffusion constant over the whole concentration range where
nematic phase is stable.

It is useful to compare our results to previous theoretical and
experimental work, especially the measurements of the diffusion
coefficient for silica coated boehmite rods~\cite{Bruggen98}. In
this work authors measure $D_{\parallel}/D_{\perp}~\approx 2$ for
monodomain nematic samples which are in coexistence with isotropic
phase. This is significantly different from
$D_{\parallel}/D_{\perp}~\approx 7.5$ for {\it fd} virus. Another
significant difference is that results on boehmite indicate that
both $D_{\parallel}$ and $D_{\perp}$ are smaller then
$D_{\mbox{\scriptsize iso}}$. In contrast to our measurements
where $D_{\parallel}$ is much larger and $D_{\perp}$ is much
smaller then $D_{\mbox{\scriptsize iso}}$.

When comparing our data to simulations of the diffusion of hard
spherocylinders and ellipsoids~\cite{Lowen99,Allen90}, one needs
to compare equivalent samples. Scaling to rod concentration where
the I-N transition takes place would be erroneous, since {\it fd}
virus is a semi-flexible rod. The semi-flexibility of the virus
drives the isotropic-nematic phase transition to higher
concentrations and it significantly decreases the order parameter
of the nematic phase in coexistence with the isotropic
phase~\cite{Tang95,Purdy03}. We choose to compare data and
simulations at the same value of the nematic order parameter which
is determined independently~\cite{Purdy03}. For {\it fd}, the
nematic order parameter is 0.65 at the I-N coexistence,
monotonically increases with increasing rod concentration and
saturates at high rod concentration. Experiment and simulation
qualitatively agree and both show a rapid increase of
$D_{\parallel}/D_{\perp}$ ratio with increasing nematic order
parameter (Fig.~\ref{diffvscon}b). We note that there is a
discrepancy between the simulations results obtain in references
~\cite{Lowen99,Allen90} which might be due to different systems
studied in these two paper.

Interestingly, simulations predict that upon increasing rod
concentration beyond I-N coexistence $D_{\parallel}$  increases
and subsequently upon approaching the smectic phase it will
decrease. The authors argues that the non-monotonic behavior of
$D_{\parallel}$ is the result of the interplay between two
effects. First, with increasing rod concentration the nematic
order parameter increases which enhances $D_{\parallel}$. Second,
with increasing rod concentration there is less free volume which
leads to decrease of $D_{\parallel}$. The author further argues
that the first effect dominates at low rod concentrations where
the nematic order parameter rapidly increases while the second
effect dominates at high rod concentrations where the nematic
order parameter is almost saturated. In contrast, both of these
effects contribute to a monotonic decrease in $D_{\perp}$ with
increasing concentration, which is observed in simulations. Due to
relatively large error in our experimental data, it is not clear
if the behavior of $D_{\parallel}$ is non-monotonic. There is an
initial hesitation, but $D_{\parallel}$ decreases over most of the
concentration range. This difference between simulations and
experiment might be because we compare experiments of
semi-flexible {\it fd} to simulations of perfectly rigid rods.
Compared to semi-flexible rods, the order parameter of rigid rods
increases much faster with increasing rod
concentration~\cite{Purdy03}.

It would be of interest to extend our measurements to rotational
diffusion in the isotropic and nematic phase. At present the rod
undergoes significant rotational diffusion during each exposure
which reduces resolution and prevents accurate determination of
the instantaneous orientation of a rod. It might be possible to
significantly reduce the exposure time by either using a more
sensitive CCD camera or a more intense laser as a illumination
source.

\section{Conclusions}
Using fluorescence microscopy we have visualized rod-like viruses
and measured the anisotropic long-time self-diffusion coefficients
in the isotropic and nematic phase. In the nematic phase the
diffusion along the director and the diffusion perpendicular to
the director decreases monotonically with increasing rod
concentration. The ratio of parallel to perpendicular diffusion
increases monotonically with increasing rod concentration. The
results compare qualitatively with simulations on hard rods with
moderate aspect ratios.

\acknowledgments Pavlik Lettinga is supported in part by
Transregio SFB TR6, "Physics of colloidal dispersions in external
fields". Zvonimir Dogic is supported by Junior Fellowship from
Rowland Institute at Harvard.

\end{document}